\pdfoutput=1

\documentclass[11pt,twoside,a4paper,cmspaper,final,collab]{cms-tdr}
\def\svnVersion{446451P}\def\cmsCernNoTag{CERN-EP-2017-310}\def\cmsCernDate{\today}\def\cmsMessage{Published in the Journal of High Energy Physics as \href{http://dx.doi.org/10.1007/JHEP02(2018)032}{\doi{10.1007/JHEP02(2018)032.}}}

\begin{document}\cmsNoteHeader{FSQ-16-005}

\hyphenation{had-ron-i-za-tion}
\hyphenation{cal-or-i-me-ter}
\hyphenation{de-vices}
\RCS$HeadURL: svn+ssh://svn.cern.ch/reps/tdr2/papers/FSQ-16-005/trunk/FSQ-16-005.tex $
\RCS$Id: FSQ-16-005.tex 446406 2018-02-16 04:39:29Z anmehta $
\providecommand{\NA}{\ensuremath{\text{---}}}

\cmsNoteHeader{FSQ-16-005}

\title{Constraints on the double-parton scattering cross section from same-sign W boson pair production in proton-proton collisions at $\sqrt{s} = 8$\TeV}

\date{\today}

\abstract{
A first search for same-sign $\PW\PW$ production via double-parton scattering is performed based on proton-proton collision data at a center-of-mass energy of 8\TeV using dimuon and electron-muon final states. The search is based on the analysis of data corresponding to an integrated luminosity of 19.7\fbinv. No significant excess of events is observed above the expected single-parton scattering yields. A 95\% confidence level upper limit of 0.32\unit{pb} is set on the inclusive cross section for same-sign $\PW\PW$ production via the double-parton scattering process. This upper limit is used to place a 95\% confidence level lower limit of 12.2\unit{mb} on the effective double-parton cross section parameter, closely related to the transverse distribution of partons in the proton. This limit on the effective cross section is consistent with previous measurements as well as with Monte Carlo event generator predictions.}

\hypersetup{
pdfauthor={CMS Collaboration},%
pdftitle={Constraints on the double-parton scattering cross section from same-sign W boson pair production in proton-proton collisions at sqrt(s) = 8 TeV},%
pdfsubject={CMS},%
pdfkeywords={CMS, physics, double-parton scattering, W boson pair production}}

\maketitle

\section{Introduction}
In proton-proton ($\Pp\Pp$) collisions at the CERN LHC, the large density of partons inside the proton at small $x$, where $x$ is the momentum fraction of the proton carried by a parton, results in a significant probability for the simultaneous occurrence of two or more parton-parton interactions within a single $\Pp\Pp$ collision~\cite{Sjostrand:1986ep}. These short-distance inelastic processes, called multiple-parton interactions (MPI), usually produce particles with relatively small transverse momenta ($\pt$) that predominantly constitute the so-called ``underlying event''. With increased parton densities at high center-of-mass energies, there is a nonnegligible probability for the production of high-$\pt$ or high-mass particles even from the second-hardest parton-parton scattering, a process known as double-parton scattering (DPS). The production cross section for a DPS process, $\sigma_\mathrm{AB}^\mathrm{DPS}$, involving two independent processes ``A'' and ``B'' with respective individual production cross sections $\sigma_\mathrm{A}$ and $\sigma_\mathrm{B}$, can be factorized as:
\begin{equation}
\sigma_\mathrm{AB}^\mathrm{DPS} =\frac{m}{2}\frac{\sigma_\mathrm{A} \sigma_\mathrm{B}}{\sigma_\text{eff}}, \label{eq:formula}
\end{equation}
where $m$ is a combinatorial factor ($m=1$ for identical and $m=2$ for different processes) and $\sigma_\text{eff}$ is an effective cross section, mainly determined by the transverse profile of partons inside the colliding hadrons and their overlap in a collision. Such a simple geometric interpretation of $\sigma_\text{eff}$ assumes negligible parton-parton correlations (in momentum, space, colour, flavour,\ldots)~\cite{dEnterria:2017yhd}, which is an assumption particularly well justified at low $x$ values where the parton densities are very large~\cite{wwDPS1}.

The measurement of the DPS cross section is important as it provides valuable information on the distribution of partons inside the proton in the transverse direction and on the correlations between them~\cite{dEnterria:2017yhd,wwDPS1,Calucci:2010wg,Rinaldi:2014ddl,Diehl:2011yj,Ceccopieri:2017oqe}. DPS also constitutes a background to searches for new physics, in rare final states with multiple heavy particles, as well as to measurements of standard model processes, such as the associated production of a Higgs and a $\PW$ or $\cPZ$ boson~\cite{Hussein:2006xr,Bandurin:2010gn}. Studies of DPS have been proposed using a variety of processes, including double Drell--Yan (DY) production~\cite{dyDPS2}, the production of same-sign $\PW$ bosons~\cite{wwDPS1}, $\PW$ or $\cPZ$ boson production in association with jets~\cite{wjetDPS1,Bansal:2016iri}, and four-jet production~\cite{jetDPS1,jetDPS2}. A number of experiments have previously measured DPS cross sections, using various final states at different collision energies~\cite{UA2,AFS,CDF1,CDF2,D0, atlas,Chatrchyan:2013xxa,Seymour1}. The magnitude of the cross section for a given DPS process depends on the value of $\sigma_\text{eff}$ and on the cross sections for the individual single-parton scattering (SPS) processes involved, according to Eq.~(\ref{eq:formula}). In the simplest approaches, $\sigma_\text{eff}$ is expected to be independent of collision energy and of the processes involved~\cite{dEnterria:2017yhd,PhysRevD.63.057901,PhysRevD.76.076006,Calucci:2010wg,Rinaldi:2014ddl}. Values of $\sigma_\text{eff}\approx 20$\unit{mb} are predicted by Monte Carlo (MC) event generators, tuned to reproduce low-$\pt$ MPI measurements~\cite{Khachatryan:2015pea}, that assume the independence of $\sigma_\text{eff}$ with respect to the scale of MPI, as defined by the momentum transfer in a given parton-parton interaction. However, the existing measurements of $\sigma_\text{eff}$ have large systematic uncertainties~\cite{Chatrchyan:2013xxa} and hence it is not possible to draw a firm conclusion about the dependence of $\sigma_\text{eff}$ on either the process or the collision energy. It is therefore important to perform further DPS cross section measurements using a variety of processes at different center-of-mass energies.

\begin{figure}[htbp]
\centering
\includegraphics[height=0.27\textheight]{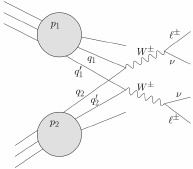}
~\includegraphics[width=0.47\textwidth]{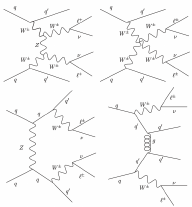}
\caption{Schematic diagrams corresponding to the production of a same-sign W boson pair via the DPS process (left) and via SPS processes (right).}
\label{fig:process}
\end{figure}

This paper presents the first measurement of the DPS process for same-sign $\PW\PW$ events in the dilepton final state using $\Pp\Pp$ collision data collected by the CMS experiment at a center-of-mass energy of $\sqrt{s}$ = 8\TeV. In the case of $\PW\PW$ production via DPS, the scale of the second hard interaction is comparable to the mass of the $\PW$ boson, which is the largest scale explored experimentally so far in DPS cross section measurements. Only same-sign $\PW\PW$ events are considered in order to suppress the contribution from the DY and SPS processes. Leptonic decays of the two $\PW$ bosons into either a pair of muons or an electron-muon pair are considered, as only these $\PW$ decay channels result in a properly-reconstructed final state that is not completely overwhelmed by background. Figure~\ref{fig:process} illustrates the production of a same-sign $\PW$ boson pair via the DPS process (left) and via a selection of leading order SPS processes (right). A set of DPS-sensitive observables is used in a multivariate analysis based on boosted decision trees (BDT) to enhance the signal sensitivity. The shape of the BDT discriminant is then used to set a limit on the cross section for same-sign $\PW\PW$ production via DPS, and subsequently on $\sigma_\text{eff}$.

This paper is organized as follows: In Section~\ref{sec:det}, a brief description of the CMS detector is presented, followed by a description of the data and the simulated samples in Section~\ref{sec:samp}. The event selection criteria, a description of the BDT, and the systematic uncertainties affecting the measurement are described in Section~\ref{sec:main}. The results are presented in Section~\ref{sec:results}, and Section~\ref{sec:sum} summarizes the studies presented here.
\section{The CMS detector}\label{sec:det}

The central feature of the CMS apparatus is a superconducting solenoid of 6\unit{m} internal diameter, providing a magnetic field of 3.8\unit{T}. Within the solenoid volume are a silicon pixel and strip tracker, a lead tungstate crystal electromagnetic calorimeter (ECAL), and a brass and scintillator hadron calorimeter (HCAL), each composed of a barrel and two endcap sections. Forward calorimeters extend the pseudorapidity ($\eta$) coverage provided by the barrel and endcap detectors. Muons are detected and measured using the gas-ionization chambers embedded in the steel flux-return yoke outside the solenoid. The first level (L1) of the CMS trigger and data acquisition systems is designed to select potentially interesting events with high efficiency~\cite{Khachatryan:2016bia}. The L1 trigger uses information collected by the calorimeters and muon detectors to select the most interesting events in less than 4\unit{$\mu$s}. The detector data are pipelined to ensure negligible deadtime up to a L1 rate of 100\unit{kHz}. After L1 triggering, data are transferred from the readout electronics of all subdetectors to the high-level trigger processor farm, where a further reduction of event rate to few hundred Hz is achieved for the purpose of data storage. A more detailed description of the CMS detector, together with a definition of the coordinate system used and the relevant kinematic variables, can be found in Ref.~\cite{Chatrchyan:2008zzk}.
\section{Data and simulated samples}\label{sec:samp}

The analyzed data correspond to an integrated luminosity of 19.7\fbinv recorded by the CMS detector during 2012 in $\Pp\Pp$ collisions at $\sqrt{s} = 8$\TeV. The decays of $\PW$ bosons into a muon or an electron (plus the corresponding neutrinos) are considered, but only the same-sign dimuon and electron-muon final states are actually used in the current analysis. These final states also include the contributions from the leptonic decay of $\tau$ leptons coming from the $\PW$ bosons. The dielectron final state is not considered because of the relatively high probability of charge misidentification for electrons, which results in this final state being overwhelmed by background from the DY process. The trigger used to select dimuon events requires the presence of a pair of muons with the leading (subleading) muon having $\pt > 17\,(8)$\GeV. The dilepton trigger, used for the online selection of the electron-muon final state, required one electron (muon) with $\pt > 17\GeV$ and one muon (electron) with $\pt > 8\GeV$. The efficiencies of the dimuon and electron-muon triggers with respect to the offline selection are 90\% and 94\%, respectively.

The simulated signal events for DPS $\PW$ boson pair production are generated using the \PYTHIA{}8 event generator (version 8.165) with the 4C tune~\cite{Sjostrand:2007gs,Corke:2010yf} to describe the underlying event processes. The contribution of $\PW$ boson pair production via SPS is removed from the signal sample. In \PYTHIA{}8, MPI are predominantly driven by the amount of overlap of the transverse matter distributions of the protons in impact parameter space~\cite{Sjostrand:1986ep}, and are interleaved with parton showering. For the tune used, the DPS cross section for (leading order) inclusive same-sign $\PW\PW$ production (including all $\PW$ boson decays) is 0.30\unit{pb}, and the corresponding effective DPS cross section amounts to $\sigma_\text{eff}=28$\unit{mb}.

Several SPS processes share the same like-sign dilepton final state as our DPS signal. All backgrounds have been studied in detail with MC simulated events as well as with data-driven estimates. The production of same-sign $\PW$ boson pairs, electroweak and strong production of $\PW$ boson pairs in association with jets ($\PW\PW$+jets), fully leptonic decays of top quark-antiquark pairs (\ttbar), DY, $\PW\cPgg^{*}$, and $\PW$/$\cPZ\PGg$ events are simulated using the \textsc{madgraph5}\xspace (version 5.1.3.30) event generator~\cite{MadGraph}. The single top quark production processes in $t$- and $s$-channels are modeled using the \POWHEG (version 1.0) event generator~\cite{Alioli:2010xd}. The $\PW\cPZ$ and $\cPZ\cPZ$ production processes are generated with the \PYTHIA{}6 event generator. All simulated samples use the CTEQ6L1~\cite{Lai:2010nw} parton density functions (PDF) set, with parton showering and hadronization performed with \PYTHIA{}6 (version 6.4.25) using the Z2* tune for the modeling of underlying event activity~\cite{Sjostrand:2006za,Z2star}. The generated MC simulations are scaled to their respective theoretical cross sections (at next-to-leading order or next-to-next-to-leading order (NNLO) accuracy, the highest order prediction available in each case)~\cite{WW-WZ,Czakon:2011xx,Kant:2014oha,Chatrchyan:2013iaa}, and multiplied by the integrated luminosity of the data sample. In addition, other background processes that result from jets being misidentified as leptons---such as single $\PW$ boson production in association with jets ($\PW$+jets), \ttbar in lepton+jets, and quantum chromodynamics (QCD) multijet production---are directly estimated from the data, as discussed in Section~\ref{bkgest}.

 The data sample analysed in this work was collected with high instantaneous luminosities which lead to additional $\Pp\Pp$ interactions (pileup) produced within the same bunch crossing. The simulated samples include the effect of pileup, with a multiplicity of pp interactions matching that from the data. The average number of measured pileup interactions per beam crossing in the 8\TeV data set is about 21. The detector response is simulated using the \GEANTfour package~\cite{Agostinelli:2002hh} and the resulting simulated events are reconstructed with the same algorithms used for the data.

\section{Experimental methods}\label{sec:main}

\subsection{Event selection}\label{sel}

A particle-flow (PF) algorithm~\cite{PF} is used for event reconstruction. The information from all subdetectors of the CMS detector is combined to reconstruct individual candidates for muons, electrons, photons, as well as charged and neutral hadrons produced in an event.

 The offline event selection criteria require the presence of at least two well reconstructed and isolated leptons with the same sign (either two muons or an electron and a muon). The leading (subleading) lepton is required to have $\pt > 20\,(10)$\GeV. The muon candidates are identified using charged-particle tracks reconstructed in the muon system that are compatible with the tracks reconstructed in the central tracking system~\cite{Chatrchyan:2012xi}. The muon candidates are required to lie within a geometrical acceptance defined by $\abs{\eta} < 2.4$. The electrons are identified using a multivariate approach based on shower shape variables, the energy sharing between the ECAL and HCAL, and the matching information provided by the tracker~\cite{Khachatryan:2015hwa}. The electrons with $\abs{\eta}<2.5$, except those falling in the transition region between the barrel and endcap of the ECAL ($1.44<\abs{\eta}<1.57$), are considered for this analysis.

A lepton isolation variable ($\mathrm{R}_\text{Iso}$)~\cite{Chatrchyan:2013iaa}, measured relative to the lepton \pt, is used to discriminate between the prompt leptons originating from a $\PW$/$\cPZ$ boson decay and those from quark and hadron decays. This variable is defined based on the sum of the transverse energies of all reconstructed particles, charged or neutral, within a cone of $\Delta R = \sqrt{\smash[b]{(\Delta\eta)^2+(\Delta\phi)^2}} < 0.3$ around the lepton direction, after subtracting the contributions from pileup and underlying event activity~\cite{Cacciari:2011ma,salam:2008} on an event-by-event basis. The value of $\mathrm{R}_\text{Iso}$ is required to be smaller than 0.12\,(0.15) for muon (electron) candidates. The two lepton candidates also need to be associated with the same primary vertex, through the requirement that the longitudinal (transverse) impact parameter of each lepton is smaller than 0.1\,(0.02)\unit{cm}.

The missing transverse momentum vector (\ptvecmiss) is defined as the projection of the negative vector sum of the momenta of all reconstructed PF objects in an event onto the plane perpendicular to the beam axis. Its magnitude is referred to as $\ptmiss$, and is corrected for anisotropic detector responses, inactive calorimeter cells, and detector misalignment. To suppress $\cPZ \to \ell^{+}\ell^{-}$ contributions, $\ptmiss$ is required to be greater than 20\GeV.

The jets are reconstructed using the anti-\kt clustering algorithm with the \FASTJET (version 2.1) package~\cite{antikt,Cacciari:2011ma} with a distance parameter of 0.5. To eliminate the jets originating from or being seeded by noisy channels in the calorimeters, a jet quality requirement, primarily based on the energy ratio between the charged and neutral hadrons, is applied~\cite{QCD-11-004JetIso}. Jet energy scale corrections~\cite{Chatrchyan:2011ds,Khachatryan:2016kdb} are used to account for the nonlinear energy response of the calorimeters and other instrumental effects. The effect of jet energy scale corrections is also propagated to $\ptmiss$.

To reduce the contributions from $\cPZ\cPZ$, $\PW \cPZ$, and $\PW\cPgg^{*}$ production processes, where the final state can have more than two leptons, events having three or more well reconstructed and isolated leptons with \pt $>$ 10\GeV are rejected. Furthermore, to reduce events from low-mass resonances, the two selected leptons are required to have an invariant mass ($m_{\ell\ell}$) greater than 20\GeV. Additionally, for the dimuon final state, $m_{\ell\ell}$ is also required to be away from the $\cPZ$ boson mass peak ($m_{\ell\ell} \notin [75, 105]\GeV$). A minimum threshold of 45\GeV on the scalar sum of the \pt of the two muons is also applied to reduce the contributions from QCD multijet events.

The main background in the electron-muon final state comes from events in which a pair of top quarks are produced and subsequently decay via their semileptonic mode $\cPqt \to \cPqb \PW$; $\PW \to \ell \PGnl$, with $\ell = \Pe, \PGm, \PGt$. The contribution from this background for the dimuon channel is found to be negligible. A b jet veto is applied in the electron-muon final state to reduce the contribution from this source. The combined secondary vertex b tagging algorithm~\cite{Chatrchyan:2012jua} is used to identify jets that are likely to originate from the hadronization of $\cPqb$ quarks. Events containing one or more b-tagged jets with $\pt >30$\GeV and $\abs{\eta} < $ 2.1 are vetoed. The b tagging efficiency is 60--80\%, while the mistag rate for light-flavored jets is about 2--3\% after the same-sign $\PW\PW$ selection criteria, given in Table~\ref{tab:selection}, have been applied.

\begin{table}
\centering
\topcaption{Event selection criteria for same-sign $\PW$ boson pair production in dimuon and electron-muon channels.}
\label{tab:selection}
\begin{tabular}{c{c}@{\hspace*{5pt}}r}
\hline
Dimuon channel                                                && Electron-muon channel    \\ \cline{1-1} \cline{3-3}
\hline
\multicolumn{3}{c}{Pair of same-sign leptons} \\
\multicolumn{3}{c}{Leading lepton $\pt > 20$\GeV} \\
\multicolumn{3}{c}{Subleading  lepton $\pt > 10$\GeV}      \\
\multicolumn{3}{c}{No third isolated and identified lepton with $\pt > 10$\GeV}      \\
\multicolumn{3}{c}{$\ptmiss> 20$\GeV}\\
\multicolumn{3}{c}{$m_{\ell\ell} >  20$\GeV}\\
\multicolumn{1}{l}{$m_{\ell\ell} \notin$ [75, 105]\GeV}                          && \NA       \\
\multicolumn{1}{l}{$|p_{\mathrm{T}_{\mu_{1}}}| + |p_{\mathrm{T}_{\mu_{2}}}| >  45$\GeV }          && \NA  \\
\multicolumn{1}{l}{\NA}                                                            && No b-tagged jet with $\pt > 30$\GeV and $\abs{\eta} < 2.1$            \\
 \hline
\end{tabular}
\end{table}

\subsection{Background evaluation}\label{bkgest}

The majority of background events originate from processes in which one or both of the leptons, coming from leptonic decays of heavy quarks or in-flight decays of light mesons, pass the event selection criteria. In the case of the electrons, overlaps of $\Pgpz \to \PGg\PGg$ decays with charged hadrons may also contaminate the sample. These lepton candidates are referred to as \textit{misidentified leptons}. Events containing one prompt and one misidentified lepton, referred to as \textit{prompt-misid.} events, mainly come from $\PW$+jets production and from semileptonic decays of top quarks. The QCD multijet events fall into the category of \textit{misid.-misid.} events, as both leptons are misidentified. A method based on control samples in the data is used to estimate the contributions of \textit{misid.-misid.} and \textit{prompt-misid.} backgrounds~\cite{Chatrchyan:2013iaa}. The method relies on a lepton misidentification rate estimated from the efficiency for a lepton-like object, passing loose lepton selection criteria of $\mathrm{R}_\text{Iso} < $ 1.0 and $\pt >10$\GeV, to also pass the complete set of lepton selection criteria described in Section~\ref{sel}. The lepton misidentification rates are measured using a control sample in the data that is enriched with misidentified leptons, and are parametrized as a function of the lepton \pt and $\eta$.

Table~\ref{tab:fakeenreg} lists the selection criteria used to construct two regions (referred to as Region 1 and Region 2) in the data that are enriched with misidentified leptons. Region 1 is used for the dimuon final state while Region 2, which additionally requires the presence of at least one b-tagged jet, is used in the electron-muon final state, since it includes a major contribution from semileptonically decaying \ttbar events. Both regions require the presence of only one loosely identified (``loose'') lepton in order to suppress $\cPZ \to \ell^{+}\ell^{-}$ contributions. Also, to further reduce the contributions from $\PW$/$\cPZ$ boson decays in the regions enriched with misidentified leptons, the transverse mass of the lepton and $\ptmiss$, $m_\mathrm{T}$($\ell,\ptmiss$), is required to be less than 20\GeV and $\ptmiss$ to be less than 20\GeV. The backgrounds with one prompt and one misidentified lepton are estimated using the \textit{tight-fail} control sample that is constructed by requiring that one of the leptons passes the loose selection criteria only, whilst the other passes the full lepton selection criteria. Similarly, another control sample with \textit{fail-fail} lepton pairs is defined in which both of the leptons pass only the loose selection criteria. Finally, the selection criteria, given in Table~\ref{tab:selection}, are applied to these samples and the resulting numbers of events are scaled using the lepton misidentification rate to estimate the contributions from \textit{ prompt-misid.} and \textit{misid.-misid.} backgrounds in the signal region.

\begin{table}[htb]
\centering
\topcaption{Control regions enriched with misidentified leptons used to extract the lepton misidentification rate. Region 1 is used for the dimuon channel. Region 2, with the additional requirement of least one b-tagged jet, is used in the electron-muon channel to reduce semileptonically decaying \ttbar events.}\label{tab:fakeenreg}
\begin{tabular}{l{c}@{\hspace*{5pt}}l}
\hline
Region 1                                    && Region 2                                                          \\ \hline 
Only one loose lepton with $\pt >  10$\GeV && Only one loose lepton with $\pt >  10$\GeV                       \\
$m_\mathrm{T}(\ell,\ptmiss) < $ 20\GeV   && $m_\mathrm{T} (\ell,\ptmiss) <  20$\GeV                         \\
$\ptmiss < 20$\GeV                         && $\ptmiss < 20$\GeV                                               \\
\NA                                          && At least one b-tagged jet with $\pt > 30$\GeV \\ 
&& and $\abs{\eta} < 2.1$ \\
\hline

\end{tabular}
\end{table}

 For the $\PW\cPgg^{*}$ background contribution, a correction factor for the simulated events is obtained from a high-purity data sample enriched with $\PW\cPgg^{*}$ events, identified by the presence of three reconstructed leptons, as described in Ref.~\cite{Chatrchyan:2013iaa}. A factor of 1.5$\pm$0.3 with respect to the predicted leading-order cross section is determined. Charged dilepton final-states from DY and \ttbar decays contribute to the background when the charge of one of the leptons is misidentified. These processes also contribute to the background if a hadronically decaying $\PGt$ lepton is misidentified as an electron or a muon and combines with a prompt lepton to form a same-sign electron-muon pair. The charge misidentification probability for electrons in the data is found to be compatible with that from the simulation; these backgrounds can therefore be estimated using the simulated samples. However, due to the limited statistical precision of the MC simulated samples, the shapes of the kinematic observables are obtained with opposite-sign electron-muon pairs in order to increase the sample sizes; all the other selection criteria given in Table~\ref{tab:selection} are applied unchanged. The resulting distributions are then normalized to the corresponding same-sign yields. The normalizations of these two backgrounds are cross-checked by constructing control regions enriched with these backgrounds. To construct a DY-enriched control region, opposite-sign pairs of electrons and muons are required to have a dilepton invariant mass that satisfies $40 < m_{\ell\ell} < 80$\GeV, and a dilepton transverse mass that satisfies $m_\mathrm{T} < 60$\GeV. For the dileptonic \ttbar decays, a control region enriched with top quark events is constructed by inverting the b jet veto criteria in the opposite-sign WW selection requirements.

The background contributions arising from lepton misidentification constitute the dominant fraction (72\%) of the total event yield after the same-sign WW selection criteria have been applied for both final states.

\subsection{Multivariate analysis}
The BDT-based framework~\cite{Hocker:2007ht} is used to discriminate between the signal and the background events, combining information from a set of kinematic variables that are sensitive to the differences between DPS WW production and the background processes. The BDT is trained using the DPS signal and the major background processes, including those originating from misidentification of leptons and diboson processes. The variables used as input for the BDT are based on energy-momentum conservation and are sensitive to the energy imbalance in the reference system of the $\PW$ boson pair.

For the dimuon channel, the following set of variables has been used for the training and testing of the BDT:
\begin{itemize}
\item{\pt of the two muons: $p_{\mathrm{T}_{1}}$, $p_{\mathrm{T}_{2}}$;}
\item{$\ptmiss$;}
\item{azimuthal angular separation between the leading/subleading muon and $\ptvecmiss$: \linebreak $\Delta\phi(\vec{p}_{\mathrm{T}_{1}}, \ptvecmiss)$ and $\Delta\phi(\vec{p}_{\mathrm{T}_{2}}, \ptvecmiss)$;}
\item{azimuthal angular separation between the two muons: $\Delta\phi(\vec{p}_{\mathrm{T}_{1}},\vec{p}_{\mathrm{T}_{2}})$;}
\item{transverse mass of the leading/subleading  muon and $\ptvecmiss$:\\
$m_\mathrm{T}(\mu_{1,2},\ptmiss) = \sqrt{\smash[b]{2p_{\mathrm{T}_{1,2}}\ptmiss(1-\cos{(\Delta\phi(\vec{p}_{\mathrm{T}_{1,2}},\ptvecmiss))})}}$;}
\item{dimuon transverse mass: $m_\mathrm{T}(\mu_{1},\mu_{2}) = {\sqrt{\smash[b]{2 p_{\mathrm{T}_{1}} p_{\mathrm{T}_{2}} (1-\cos{(\Delta\phi(\vec{p}_{\mathrm{T}_{1}},\vec{p}_{\mathrm{T}_{2}}))})}}}$.}

\end{itemize}

For the electron-muon channel, the BDT variables include:
\begin{itemize}
\item{\pt of the two leptons: $p_{\mathrm{T}_{1}}$, $p_{\mathrm{T}_{2}}$;}
\item{vector sum of the \pt of the two leptons: $\vec{p}_{\mathrm{T}_{12}} = \vec{p}_{\mathrm{T}_{1}} + \vec{p}_{\mathrm{T}_{2}}$;}
\item{$\ptmiss$;}
\item{pseudorapidity separation between the two leptons: $\Delta\eta(\ell_{1},\ell_{2})$;}
\item{azimuthal angular separation between the subleading lepton and $\ptvecmiss$: $\Delta\phi(\vec{p}_{\mathrm{T}_{2}},\ptvecmiss)$;}
\item{azimuthal angular separation between the two leptons: $\Delta\phi(\vec{p}_{\mathrm{T}_{1}},\vec{p}_{\mathrm{T}_{2}})$;}
\item{azimuthal angular separation between the resultant direction of the dilepton system and $\ptvecmiss$: $\Delta\phi(\vec{p}_{\mathrm{T}_{12}},\ptvecmiss)$.}
\end{itemize}

\begin{figure}[htbp]
\centering
\includegraphics[width=0.45\textwidth]{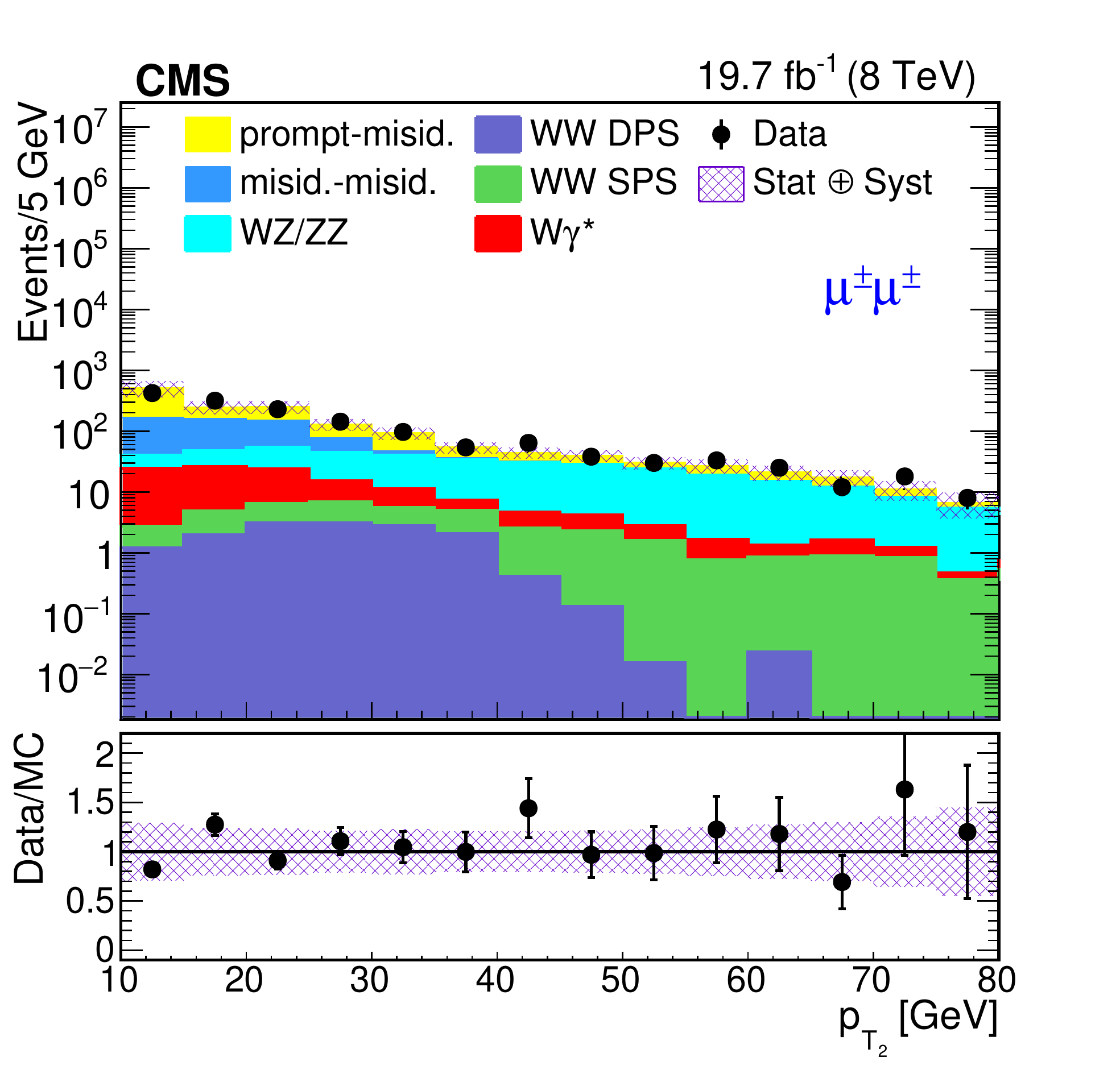}
\includegraphics[width=0.45\textwidth]{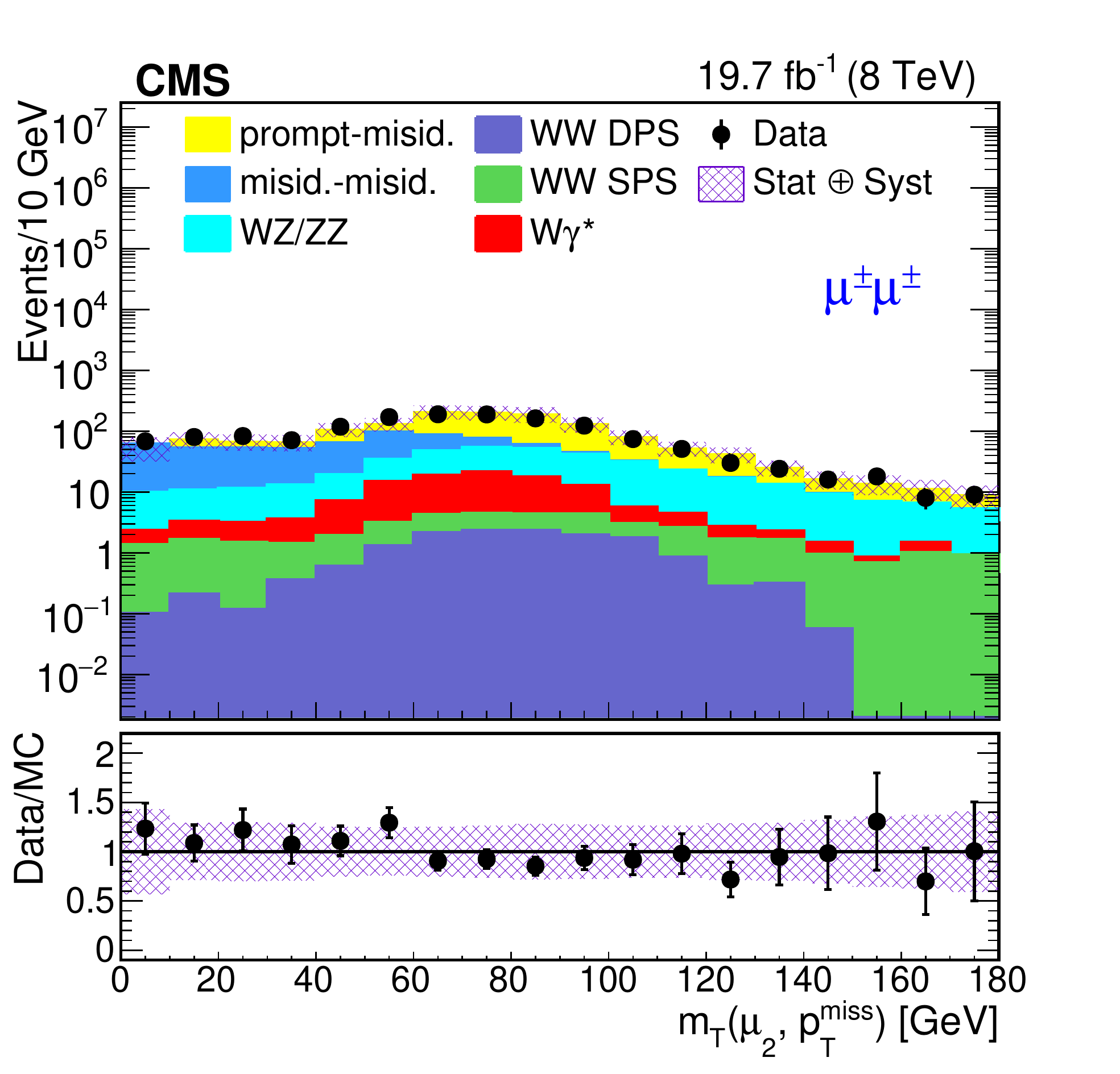}\\
\includegraphics[width=0.45\textwidth]{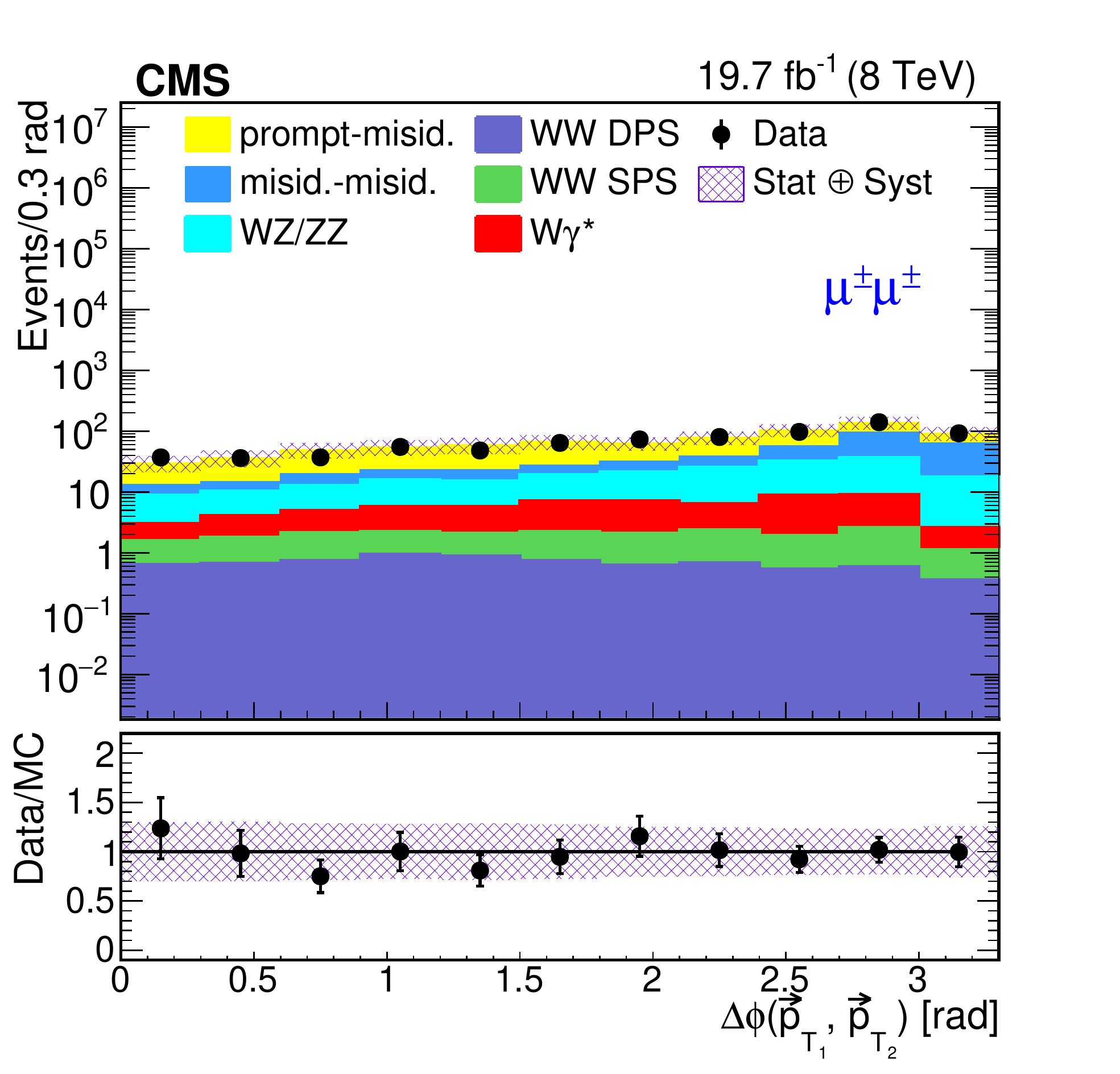}
\includegraphics[width=0.45\textwidth]{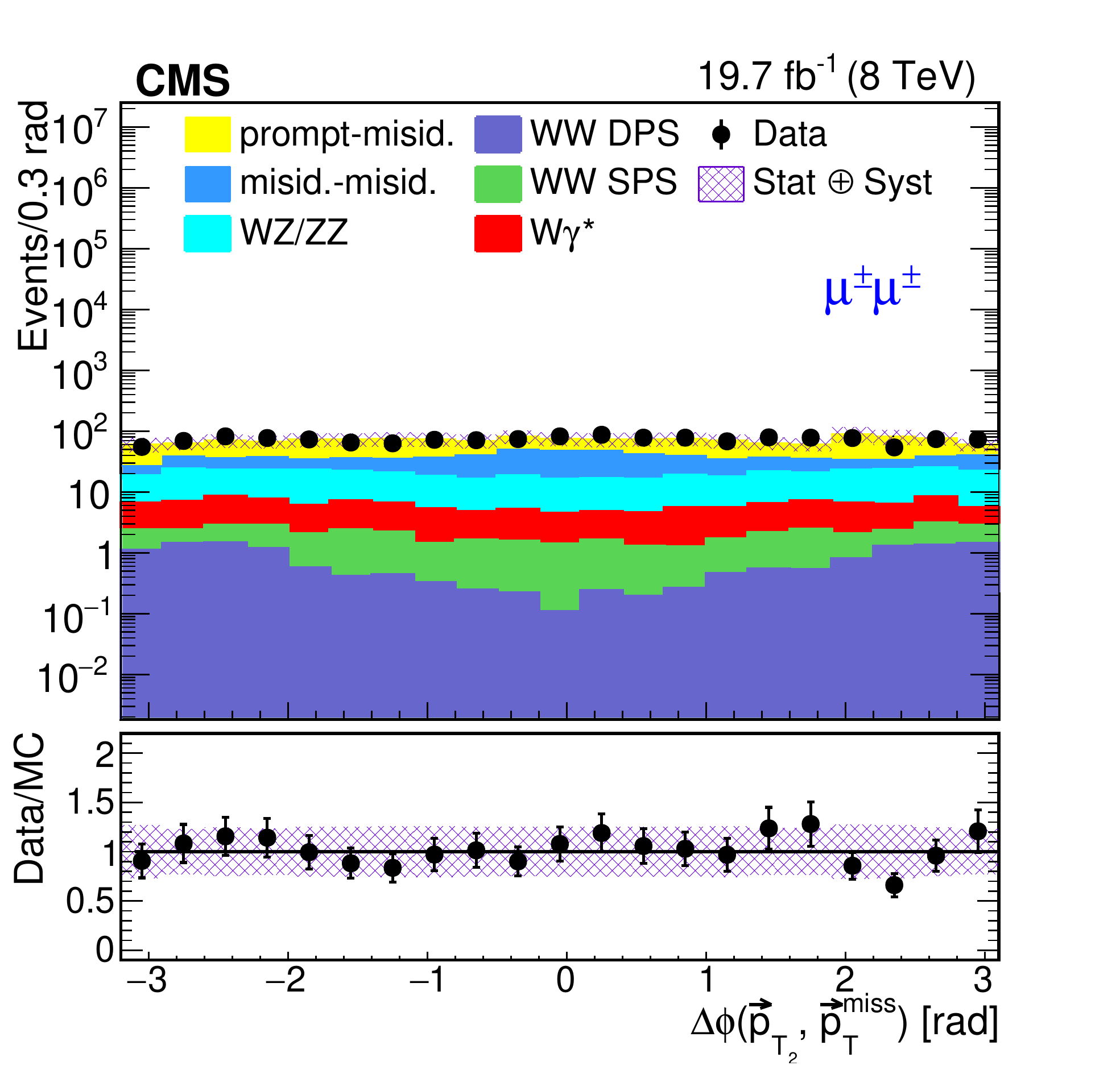}
\caption{Distributions of the $p_{\mathrm{T}_{2}}$ (top-left), $m_\mathrm{T}(\mu_{2},\ptmiss)$ (top-right), $\Delta\phi(\vec{p}_{\mathrm{T}_{1}},\vec{p}_{\mathrm{T}_{2}})$ (bottom-left), and $\Delta\phi(\vec{p}_{\mathrm{T}_{2}}, \ptvecmiss)$ (bottom-right) variables for the dimuon channel, after the same-sign $\PW\PW$ selection criteria have been applied. The data are represented by the black dots and the shaded histograms represent the predicted signal and background processes normalized according to the estimated cross sections and the luminosity. For each individual distribution, the bottom panels show the ratio of the number of events observed in the data to that predicted by the simulation, along with the associated statistical uncertainty. The hatched bands in all cases represent the sum of the systematic and statistical uncertainties of the simulated samples, added in quadrature.}\label{fig:dimuvars}
\end{figure}

\begin{figure}[htb]
\centering
\includegraphics[width=0.45\textwidth]{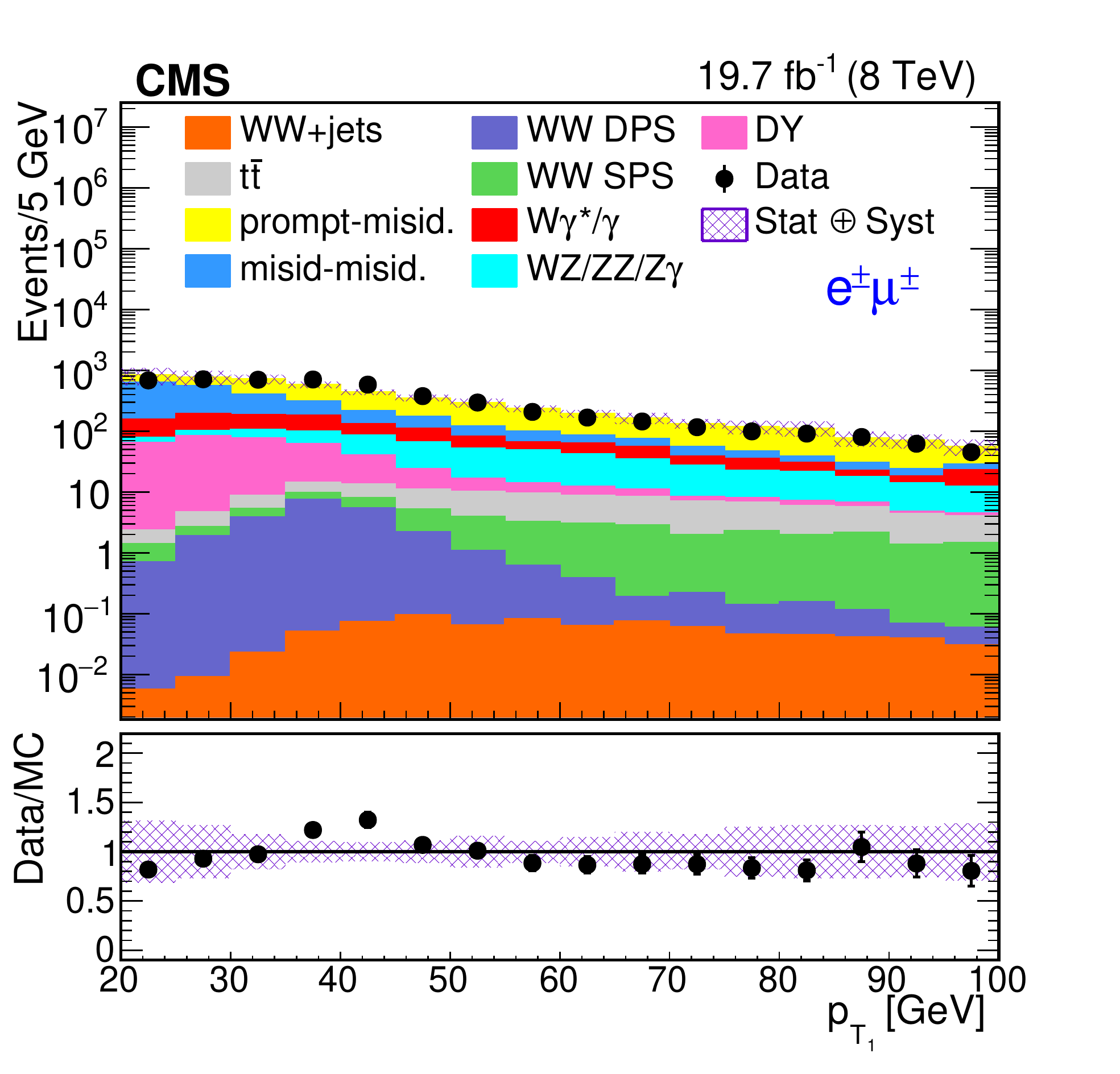}
\includegraphics[width=0.45\textwidth]{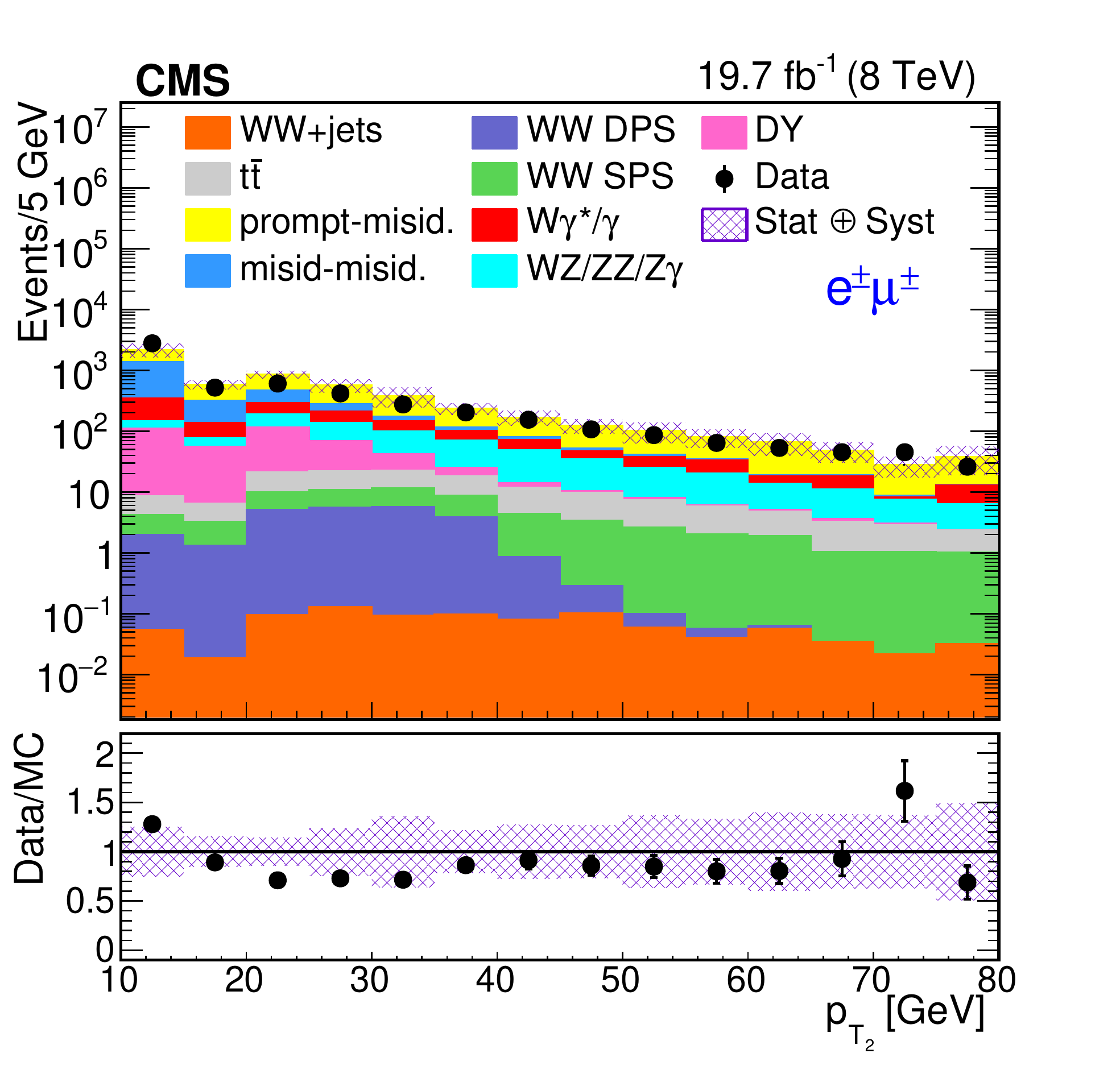}
\includegraphics[width=0.45\textwidth]{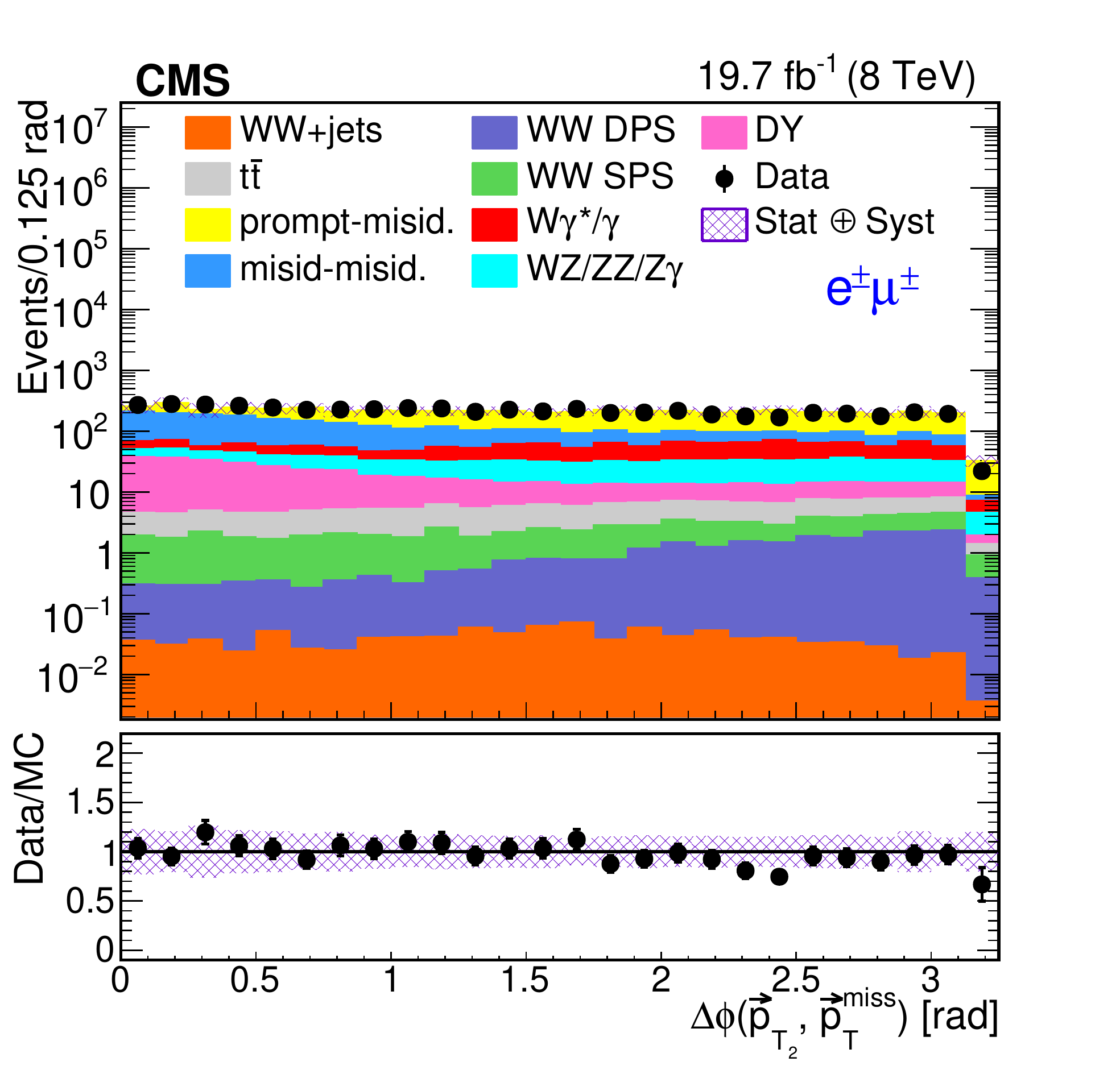}
\includegraphics[width=0.45\textwidth]{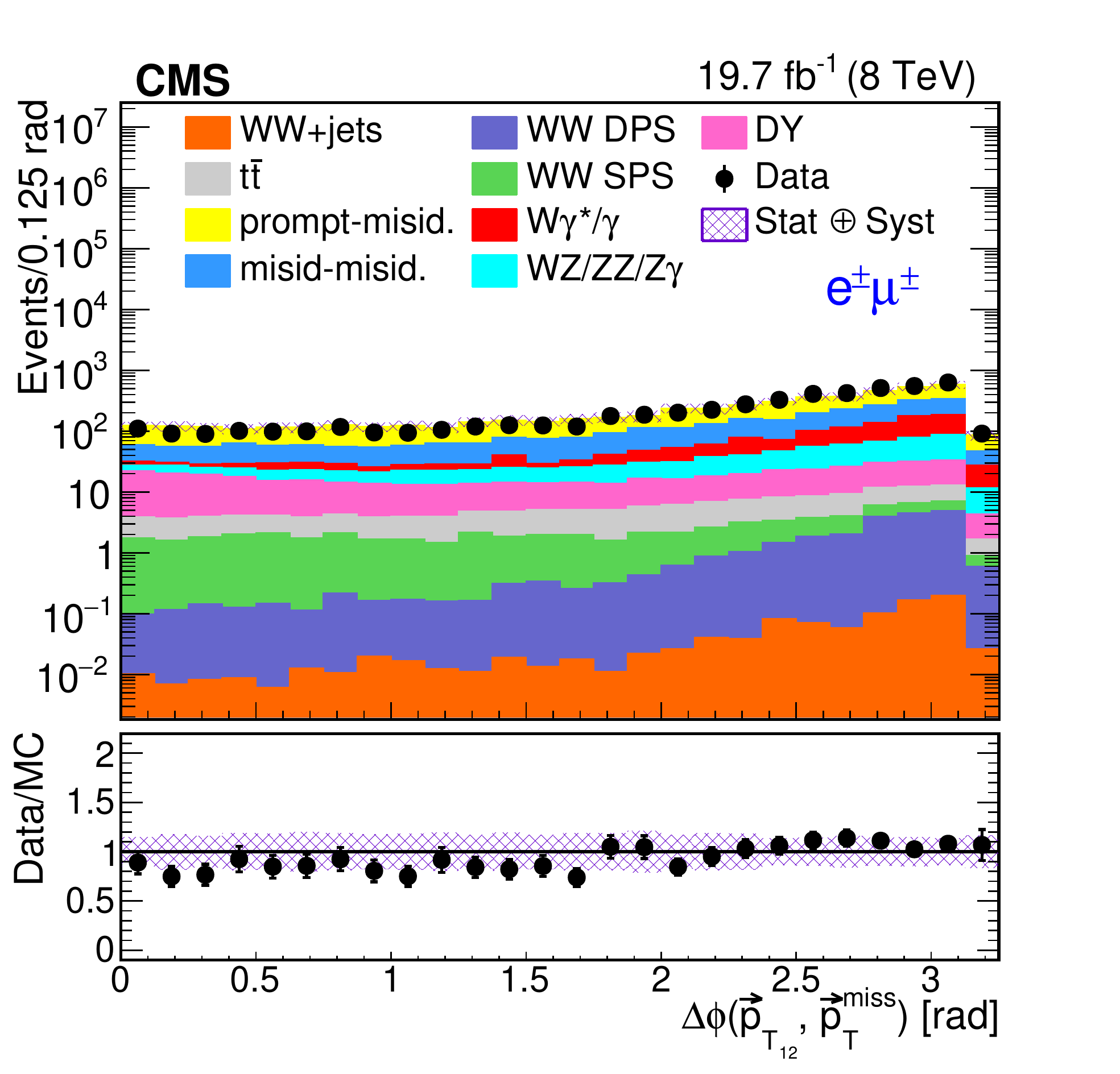}
\caption{Distributions of the $p_{\mathrm{T}_{1}}$ (top-left), $p_{\mathrm{T}_{2}}$ (top-right), $\Delta\phi(\vec{p}_{\mathrm{T}_{2}},\ptvecmiss)$ (bottom-left), and $\Delta\phi(\vec{p}_{\mathrm{T}_{12}},\ptvecmiss)$ (bottom-right) variables for the electron-muon channel, after the same-sign $\PW\PW$ selection criteria have been applied. Symbols and patterns are the same as in Fig.~\ref{fig:dimuvars}.}\label{fig:emuvars}
\end{figure}

These sets of variables have been selected based on their power to discriminate between the signal and background processes. Figures~\ref{fig:dimuvars} and \ref{fig:emuvars} compare the data to the signal and background predictions for the most sensitive of the input variables for the dimuon and electron-muon final states, respectively, after applying the same-sign WW selection criteria. Overall, the data and simulation are found to be consistent within the uncertainties.
The BDT discriminant after the full event selection has been applied is used to extract the limits on the DPS cross section and $\sigma_\text{eff}$ using statistical analysis techniques.

\subsection{Systematic uncertainties}\label{systuncert}

The systematic uncertainties in this analysis arise from the background estimation techniques, experimental measurements, and theoretical predictions.

The dominant source of systematic uncertainty is associated with the method adopted for the estimation of \textit{misid.-misid.} and \textit{prompt-misid.} backgrounds, and with the definition of the control sample used to obtain the lepton misidentification rate.

To estimate the effects of the jet \pt spectra and jet flavor on the lepton misidentification rate, these backgrounds are estimated by changing the definition of the misidentified lepton-enriched region. The observed differences in the estimated event yields and in the shapes of the kinematic observables, for the different definitions of the control samples, are taken as the systematic uncertainty.
For the dimuon channel, the lepton misidentification rate is recalculated by requiring the presence of a jet with $\pt > 25\GeV$ in addition to the nominal selection criteria for Region 1. To estimate the effect of jet flavor, the lepton misidentification rate is measured using the QCD multijet simulated sample and applied to the $\PW$+jets simulated sample.

For the electron-muon channel, these backgrounds are recalculated after removing the requirement of the presence of a b-tagged jet in the definition of the misidentified lepton-enriched region. The effect of statistical fluctuations on the lepton misidentification rate is also considered when calculating the final background yields. The systematic uncertainty arising from this background estimation method results in a 40\% variation in the \textit{misid.-misid.} event yields for both final states, and in the \textit{prompt-misid.} event yield for the dimuon channel. For the electron-muon channel this systematic uncertainty results in a 20\% to 40\% variation of the yield of \textit{prompt-misid.} events, depending on the shape of the kinematic observable being considered.

The uncertainty on the yields of the various simulated samples from pileup mismodeling is evaluated to be 4--5\%. This is determined by varying the inelastic $\Pp\Pp$ cross section, which is used to estimate the pileup contribution in data, from its central value within its $\pm5$\% uncertainty. The measurements are also affected by the uncertainty on the integrated luminosity calibration, and an uncertainty of 2.6\%~\cite{LumiUncert} is assigned to the simulated samples to account for this.

The trigger and lepton identification efficiencies in the data and simulation are measured using the ``tag-and-probe'' method~\cite{Chatrchyan:2013iaa}. The ratio of the efficiencies obtained from the data and simulation is used to scale the selection efficiency in the simulated samples. The uncertainty on this scale factor for the trigger efficiency is of the order of 1\% and is also applied to all the simulated samples. The systematic uncertainty associated with the lepton identification efficiency (1\% for muons and 4\% for electrons) is applied to all simulated samples. The lepton momentum scale has uncertainties due to detector misalignment~\cite{Chatrchyan:2013iaa}. For the muons, a momentum scale uncertainty of 1\%, independent of its $\eta$, is assigned. A momentum scale uncertainty of 2\% is assigned for electrons in the barrel, and 4\% for electrons in the endcaps of the ECAL. The lepton momentum scale affects the final predicted yields by 1--2\% in each channel. The effects of the jet energy scale uncertainty and the jet energy resolution are evaluated by shifting the $\pt$ of the leptons and the jets by their respective uncertainties, with the effect being propagated to \ptvecmiss~\cite{Chatrchyan:2011ds,MET,Khachatryan:2016kdb}. These uncertainties cause the predicted event yields to vary by 2--4\% for the dimuon and by 5\% for the electron-muon channels, respectively.

A scale factor is applied to the simulation to correct for different b jet tagging efficiencies and mistag rates measured in the data~\cite{CMS-PAS-BTV-13-001}. This correction is applied by reweighting all the simulated samples on an event-by-event basis, where the weight depends on the flavor and kinematics of the jets. This results in an uncertainty of 4\% on the b jet dominated background and less than 1\% for other background processes. It should be noted that this particular source of systematic uncertainty affects the electron-muon channel only.

To check the normalization of the DY background for the electron-muon channel, a DY-enriched control region is constructed from the data, as defined in Section~\ref{bkgest}. A normalization uncertainty of 10\% is derived for the DY background by looking at the ratio of the data to simulation in this control region.

For the $\PW\PGg$ and $\PW\cPgg^{*}$ backgrounds, a 30\% uncertainty is derived for the normalization factor for both of the final states. The effects of varying the PDFs and the value of $\alpha_\mathrm{S}$, as well as the effect of higher-order corrections, are estimated using the PDF4LHC prescription~\cite{Alekhin:2011sk,Botje:2011sn}.

\section{Results}
\label{sec:results}
The expected and observed upper limits at 95\% confidence level (CL) on the cross section for inclusive same-sign $\PW\PW$ production via DPS have been extracted. The statistical interpretation of the results is performed using an asymptotic approximation of the CL$_\mathrm{s}$ method~\cite{cls1,Cowan:2010js,cls4}. These limits are estimated by fitting the shape of the BDT discriminant, using the methodology developed by the ATLAS and CMS Collaborations~\cite{cls2}. A log-normal probability distribution function is assumed for the nuisance parameters that affect the event yields of the signal and various background contributions. Systematic uncertainties affecting the shape of the BDT discriminant are assumed to have a Gaussian probability distribution function. A binned maximum likelihood fit is performed on the selected events while the systematic uncertainties are included in the fit as nuisance parameters and are profiled during the minimization~\cite{cls2}.

While performing the combination of the results from the two final states, the systematic uncertainties arising from theoretical predictions or from the background estimation techniques are taken to be fully correlated across the two final states, while no correlation is assumed for uncertainties of statistical origin. The uncertainty associated with the absolute scale of the integrated luminosity and the effects of pileup are correlated across the two final states. Experimental uncertainties on the lepton selection and trigger efficiencies for the same kind of physics objects are assumed to be correlated. Theoretical uncertainties on the production cross sections for each process are correlated across the two final states. However, the uncertainties on different processes are assumed to be independent.

\begin{figure}[htbp]
\centering
\includegraphics[width=0.45\textwidth]{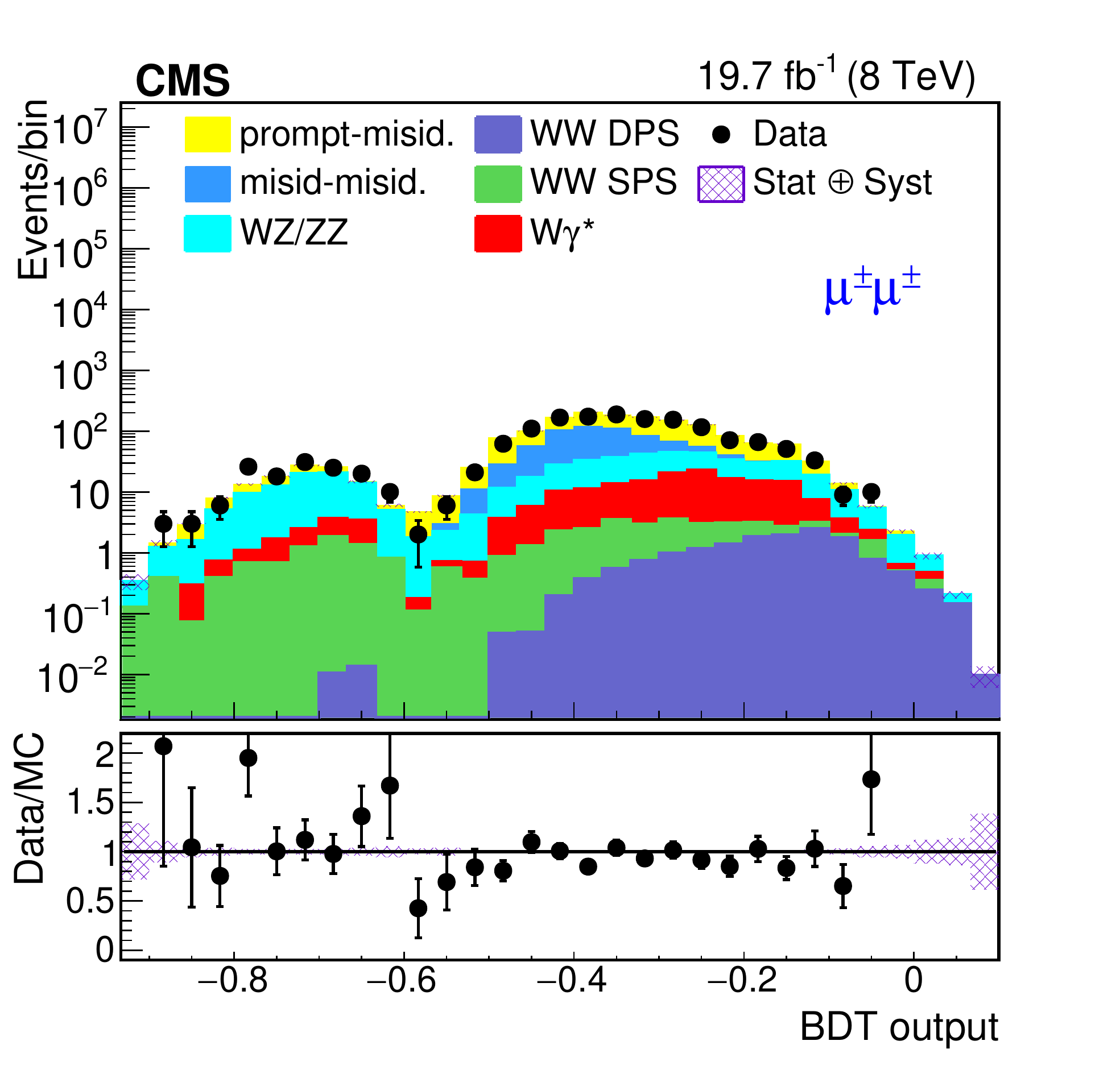}
\includegraphics[width=0.45\textwidth]{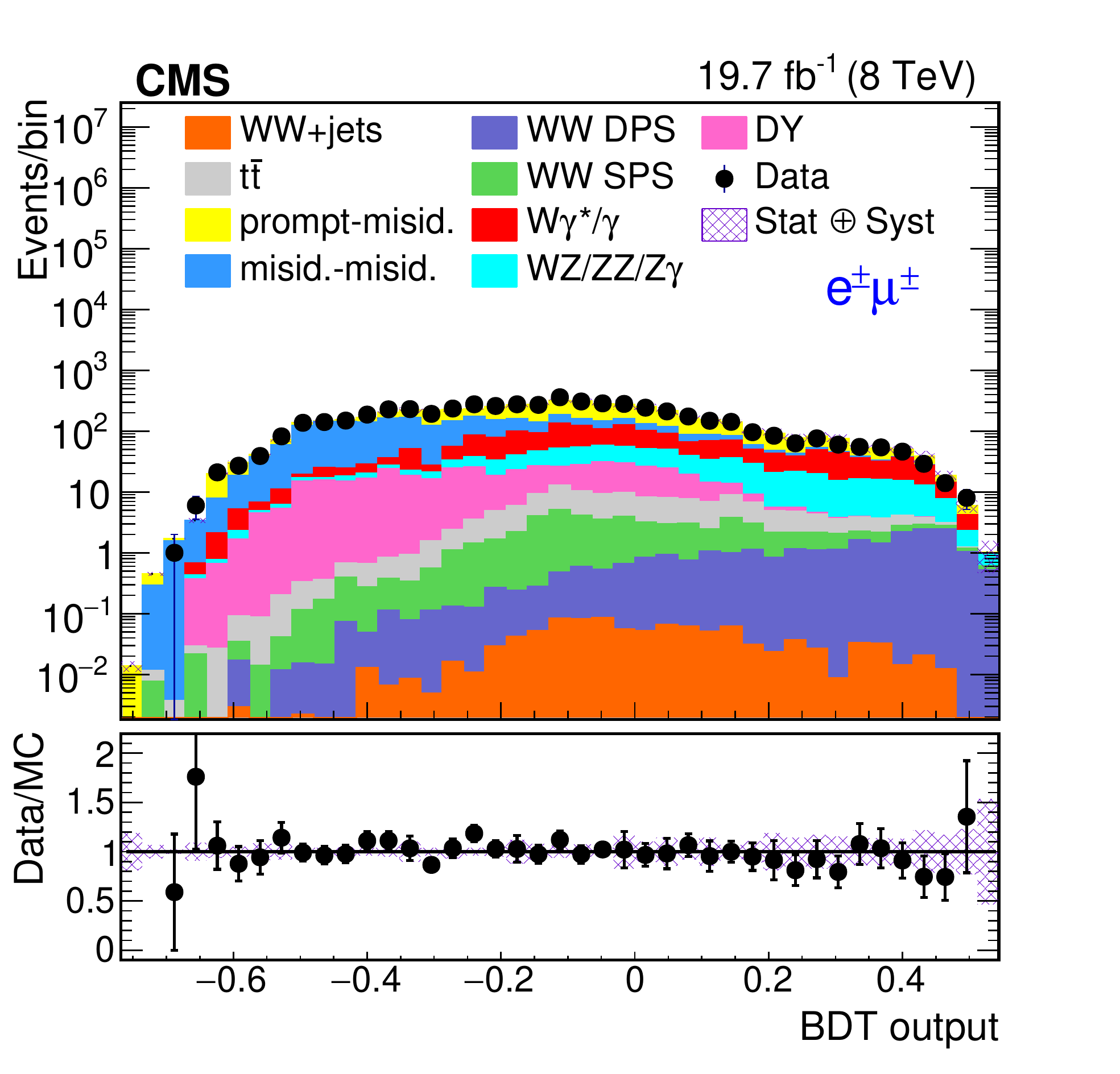}
\caption{Distribution of the BDT discriminant, for the dimuon channel (left) and for the electron-muon channel (right). The data are represented by the black dots and the shaded histograms represent the pre-fit signal and post-fit background processes. The bottom panels show the ratio of data to the sum of all signal and background contributions. The hatched bands represent the post-fit uncertainty, which includes both the statistical and systematic components.}\label{fig:postfitBDT}
\end{figure}

Figure~\ref{fig:postfitBDT} shows the distributions of the BDT discriminant having post-fit contributions for the backgrounds and pre-fit ones for the signal, for the dimuon and electron-muon final states with the corresponding uncertainty bands (shown as hatched bands). The expected and observed 95\% CL limits on the cross section for same-sign $\PW\PW$ production via DPS ($\sigma^\mathrm{DPS}_{\PW^{\pm}\PW^{\pm}}$) are summarized in Table~\ref{tab:ss_limit}.

\begin {table}[htb]
\topcaption{Expected and observed 95\% CL limits on the cross section for inclusive same-sign $\PW\PW$ production via DPS for the dimuon and electron-muon channels along with their combination.}\label{tab:ss_limit}
\centering
\begin{tabular}{lccc}
\hline
95\%  CL               & Dimuon & Electron-muon & Combined \\
\hline
Expected                & 0.67\unit{pb}         & 0.78\unit{pb}              &  0.48\unit{pb}        \\
Expected $\pm 1\sigma$   & [0.46, 1.00]\unit{pb}  & [0.52, 1.16]\unit{pb}        &  [0.33, 0.72]\unit{pb}  \\
Expected $\pm 2\sigma$   & [0.34, 1.45]\unit{pb} & [0.37, 1.71]\unit{pb}        &  [0.24, 1.04]\unit{pb}  \\[2ex]
Observed          &  0.72\unit{pb}        & 0.64\unit{pb}               &  0.32\unit{pb}       \\
\hline
\end{tabular}
\end{table}

The expected value of the DPS cross section derived with the factorization formula given by Eq.~({\ref{eq:formula}}) is $\sigma^\mathrm{DPS}_{\PW^{\pm}\PW^{\pm}}=0.18\pm0.06$\unit{pb}, as obtained for the effective cross section $\sigma_\text{eff}=20.7\pm6.6$\unit{mb} measured in the $\PW$+2 jets final state at 7\TeV~\cite{Chatrchyan:2013xxa}, and the single-parton NNLO cross sections of $\sigma_{\PW^{+}}=72.1\pm2.5$\unit{nb} and $\sigma_{\PW^{-}}=50.8\pm1.9$\unit{nb}~\cite{Gavin:2012sy} combined.

\begin{figure}[htbp]
\centering
\includegraphics[width=0.75\textwidth]{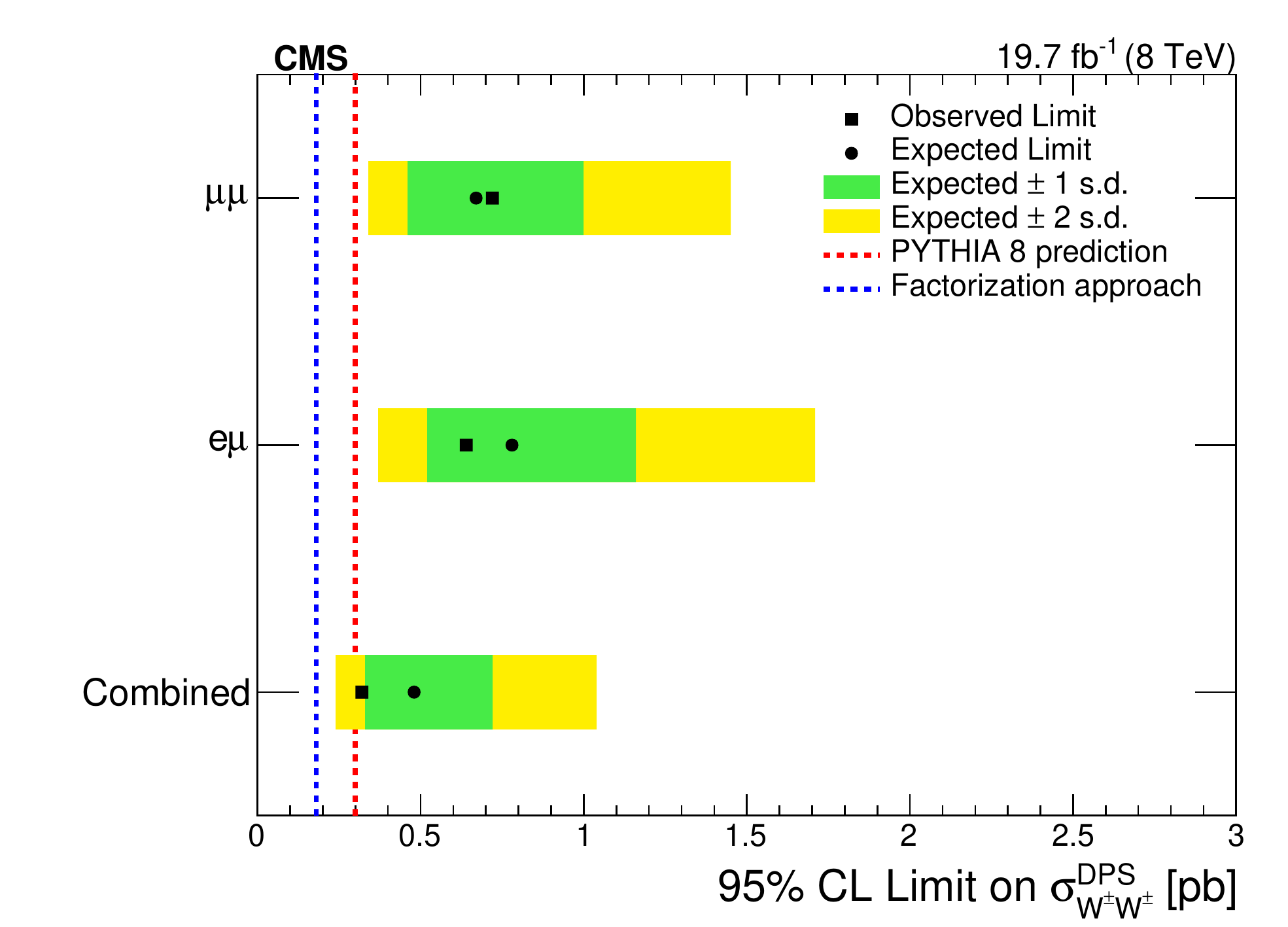}
\caption{Expected and observed 95\% CL upper limits on the same-sign $\sigma^\mathrm{DPS}_{\PW^{\pm}\PW^{\pm}}$ for the dimuon and electron-muon final states, along with their combination. The predicted values of $\sigma^\mathrm{DPS}_{\PW^{\pm}\PW^{\pm}}$ from \PYTHIA{}8 and from the factorization approach~\cite{Chatrchyan:2013xxa} are also shown.}\label{fig:limit_BDT}
\end{figure}

Figure~\ref{fig:limit_BDT} provides a summary of the sensitivity of the BDT-based analysis for the different final states. The expected value of same-sign $\sigma^\mathrm{DPS}_{\PW^{\pm}\PW^{\pm}}$ taken from \textsc{pythia8} is shown as a red line, while that extracted using the factorization approach is represented by a blue line. The observed and expected limits are consistent within the statistical fluctuations since the observed limits are within the green (68\%) or yellow (95\%) bands of the expected limit values. The observed limits for the combined analysis are more stringent than the limits from the individual final states.

Assuming the two scatterings to be independent, a limit can be placed on $\sigma_\text{eff}$ using Eq.~(\ref{eq:formula}) together with the SPS $\sigma_{\PW^{+}}$ and $\sigma_{\PW^{-}}$ cross section values at NNLO. A lower 95\% CL limit on $\sigma_\text{eff}$ can be calculated as:

\begin{equation*}
\sigma_\text{eff} > \frac{ \sigma_{\PW^{+}}^{2} + \sigma_{\PW^{-}}^{2}} {2\,\sigma^\mathrm{DPS}_{\PW^{\pm}\PW^{\pm}}} =12.2\unit{mb}.
\end{equation*}

The obtained lower limit on $\sigma_\text{eff}$ is compatible with the values of $\sigma_\text{eff}\approx$10--20\unit{mb} obtained from measurements at different center-of-mass energies using a variety of processes~\cite{Chatrchyan:2013xxa}.

\section{Summary}\label{sec:sum}
A first search for same-sign $\PW$ boson pair production via double-parton scattering (DPS) in $\Pp\Pp$ collisions at a center-of-mass energy of 8\TeV has been presented. The analyzed data were collected by the CMS detector at the LHC during 2012 and correspond to an integrated luminosity of 19.7\fbinv. The results presented here are based on the analysis of events containing two same-sign $\PW$ bosons decaying into either same-sign muon-muon or electron-muon pairs. Several kinematic observables have been studied to identify those that can better discriminate between DPS and the single-parton scattering (SPS) backgrounds. These observables with discriminating power are used as an input to a multivariate analysis based on boosted decision trees. No excess over the expected contributions from SPS processes is observed. A 95\% confidence level (CL) upper limit of 0.32\unit{pb} is placed on the inclusive cross section for same-sign $\PW\PW$ production via DPS. A corresponding 95\% CL lower limit of 12.2\unit{mb} on the effective double-parton cross section is also derived, compatible with previous measurements as well as with Monte Carlo event generator expectations.

\begin{acknowledgments}
We congratulate our colleagues in the CERN accelerator departments for the excellent performance of the LHC and thank the technical and administrative staffs at CERN and at other CMS institutes for their contributions to the success of the CMS effort. In addition, we gratefully acknowledge the computing centres and personnel of the Worldwide LHC Computing Grid for delivering so effectively the computing infrastructure essential to our analyses. Finally, we acknowledge the enduring support for the construction and operation of the LHC and the CMS detector provided by the following funding agencies: BMWFW and FWF (Austria); FNRS and FWO (Belgium); CNPq, CAPES, FAPERJ, and FAPESP (Brazil); MES (Bulgaria); CERN; CAS, MoST, and NSFC (China); COLCIENCIAS (Colombia); MSES and CSF (Croatia); RPF (Cyprus); SENESCYT (Ecuador); MoER, ERC IUT, and ERDF (Estonia); Academy of Finland, MEC, and HIP (Finland); CEA and CNRS/IN2P3 (France); BMBF, DFG, and HGF (Germany); GSRT (Greece); OTKA and NIH (Hungary); DAE and DST (India); IPM (Iran); SFI (Ireland); INFN (Italy); MSIP and NRF (Republic of Korea); LAS (Lithuania); MOE and UM (Malaysia); BUAP, CINVESTAV, CONACYT, LNS, SEP, and UASLP-FAI (Mexico); MBIE (New Zealand); PAEC (Pakistan); MSHE and NSC (Poland); FCT (Portugal); JINR (Dubna); MON, RosAtom, RAS, RFBR and RAEP (Russia); MESTD (Serbia); SEIDI, CPAN, PCTI and FEDER (Spain); Swiss Funding Agencies (Switzerland); MST (Taipei); ThEPCenter, IPST, STAR, and NSTDA (Thailand); TUBITAK and TAEK (Turkey); NASU and SFFR (Ukraine); STFC (United Kingdom); DOE and NSF (USA).

\hyphenation{Rachada-pisek} Individuals have received support from the Marie-Curie programme and the European Research Council and Horizon 2020 Grant, contract No. 675440 (European Union); the Leventis Foundation; the A. P. Sloan Foundation; the Alexander von Humboldt Foundation; the Belgian Federal Science Policy Office; the Fonds pour la Formation \`a la Recherche dans l'Industrie et dans l'Agriculture (FRIA-Belgium); the Agentschap voor Innovatie door Wetenschap en Technologie (IWT-Belgium); the Ministry of Education, Youth and Sports (MEYS) of the Czech Republic; the Council of Science and Industrial Research, India; the HOMING PLUS programme of the Foundation for Polish Science, cofinanced from European Union, Regional Development Fund, the Mobility Plus programme of the Ministry of Science and Higher Education, the National Science Center (Poland), contracts Harmonia 2014/14/M/ST2/00428, Opus 2014/13/B/ST2/02543, 2014/15/B/ST2/03998, and 2015/19/B/ST2/02861, Sonata-bis 2012/07/E/ST2/01406; the National Priorities Research Program by Qatar National Research Fund; the Programa Severo Ochoa del Principado de Asturias; the Thalis and Aristeia programmes cofinanced by EU-ESF and the Greek NSRF; the Rachadapisek Sompot Fund for Postdoctoral Fellowship, Chulalongkorn University and the Chulalongkorn Academic into Its 2nd Century Project Advancement Project (Thailand); the Welch Foundation, contract C-1845; and the Weston Havens Foundation (USA).
\end{acknowledgments}

\bibliography{auto_generated}

\cleardoublepage \appendix\section{The CMS Collaboration \label{app:collab}}\begin{sloppypar}\hyphenpenalty=5000\widowpenalty=500\clubpenalty=5000\textbf{Yerevan Physics Institute,  Yerevan,  Armenia}\\*[0pt]
A.M.~Sirunyan, A.~Tumasyan
\vskip\cmsinstskip
\textbf{Institut f\"{u}r Hochenergiephysik,  Wien,  Austria}\\*[0pt]
W.~Adam, F.~Ambrogi, E.~Asilar, T.~Bergauer, J.~Brandstetter, E.~Brondolin, M.~Dragicevic, J.~Er\"{o}, M.~Flechl, M.~Friedl, R.~Fr\"{u}hwirth\cmsAuthorMark{1}, V.M.~Ghete, J.~Grossmann, J.~Hrubec, M.~Jeitler\cmsAuthorMark{1}, A.~K\"{o}nig, N.~Krammer, I.~Kr\"{a}tschmer, D.~Liko, T.~Madlener, I.~Mikulec, E.~Pree, D.~Rabady, N.~Rad, H.~Rohringer, J.~Schieck\cmsAuthorMark{1}, R.~Sch\"{o}fbeck, M.~Spanring, D.~Spitzbart, W.~Waltenberger, J.~Wittmann, C.-E.~Wulz\cmsAuthorMark{1}, M.~Zarucki
% ...
\vskip\cmsinstskip
% ...
1:~~Also at Vienna University of Technology, Vienna, Austria\\
% ...

\end{sloppypar}
\end{document}